\documentclass[12pt, a4paper]{article}
\usepackage{epsf}
\usepackage{cite}
\usepackage{amsmath,amssymb}
\input{colordvi.tex}
\usepackage[usenames,dvipsnames]{color}
\usepackage{graphicx}
\usepackage{ascmac}
\usepackage{hyperref}

\begin{document}

\begin{titlepage}

\begin{center}
\hfill TU-1199\\
\hfill KEK-QUP-2023-0015
\vskip .5in

{\Large \bf
Nanohertz gravitational waves from cosmic strings\\ \vspace{2.5mm} 
and dark photon dark matter
}

\vskip .5in

{\large
Naoya Kitajima$^{(a,b)}$ and
Kazunori Nakayama$^{(b,c)}$
}

\vskip 0.5in

$^{(a)}${\em 
Frontier Research Institute for Interdisciplinary Sciences, Tohoku University, Sendai 980-8578, Japan
}

\vskip 0.2in

$^{(b)}${\em 
Department of Physics, Tohoku University, Sendai 980-8578, Japan
}

\vskip 0.2in

$^{(c)}${\em 
International Center for Quantum-field Measurement Systems for Studies of the Universe and Particles (QUP), KEK, 1-1 Oho, Tsukuba, Ibaraki 305-0801, Japan
}

\end{center}
\vskip .5in

\begin{abstract}
The recent observations by pulsar timing array (PTA) experiments suggest the existence of stochastic gravitational wave background in the nano-Hz range. It can be a hint for the new physics and cosmic string is one of the promising candidate. In this paper, we study the implication of the PTA result for cosmic strings and dark photon dark matter produced by the decay of cosmic string loops. It can simultaneously explain the PTA result and present dark matter abundance for the dark photon mass $m \sim 10^{-6}$--$10^{-4}\,{\rm eV}$. Implications for the gravitational wave detection with multi-frequency bands are also discussed.
\end{abstract}

\end{titlepage}


Recently pulsar timing array (PTA) collaborations, including the North American Nanohertz Observatory for Gravitational Waves (NANOGrav)~\cite{NANOGrav:2023gor}, Eulopean Pulsar Timing Array (EPTA)~\cite{Antoniadis:2023ott}, Parkes Pulsar Timing Array (PPTA)~\cite{Reardon:2023gzh} and Chiniese Pulsar Timing Array (CPTA)~\cite{Xu:2023wog}, reported a detection of stochastic gravitational waves (GW) around the frequency of $\sim 10^{-8}$\,Hz at the level of GW energy density parameter $\Omega_{\rm GW} \sim 10^{-9}$. It may be explained by the super massive black hole binaries, while interpretations in terms of new physics phenomena are also possible~\cite{NANOGrav:2023hvm}. In this letter we consider the cosmic strings that are source of the dark photon dark matter, as the origin of stochastic GWs detected by the NANOGrav.\footnote{
	See Refs.~\cite{Ellis:2020ena,Blasi:2020mfx,Buchmuller:2020lbh} for the cosmic string explanation of the NANOGrav signal reported in 2020~\cite{NANOGrav:2020bcs}. 
}
The possibility of cosmic strings have been already discussed in detail in Ref.~\cite{NANOGrav:2023hvm}, where it has been concluded that the stable cosmic strings are not suitable for explaining the PTA data, since the spectrum is too soft around the PTA frequency range. 
We point out that, for the case of cosmic strings in association with the light dark photon dark matter, the spectrum fits well with the PTA result.

There are many production mechanisms of dark photon dark matter, such as the gravitational production during inflation or reheating~\cite{Graham:2015rva,Ema:2019yrd,Ahmed:2020fhc,Kolb:2020fwh,Sato:2022jya,Redi:2022zkt,Nakai:2022dni,Tang:2017hvq,Garny:2017kha}, resonant production through the oscillation of the axion or dark-Higgs~\cite{Agrawal:2018vin,Co:2018lka,Bastero-Gil:2018uel,Dror:2018pdh,Nakayama:2021avl}, production from the cosmic string network~\cite{Long:2019lwl,Kitajima:2022lre}, misalignment production during inflation~\cite{Nakayama:2019rhg,Nakayama:2020rka,Kitajima:2023fun}.
Among them, in this letter we focus on the production from the cosmic strings~\cite{Long:2019lwl,Kitajima:2022lre}.
We consider the following Abelian-Higgs model with the dark U(1) gauge symmetry:
\begin{align}
	\mathcal L = -\frac{1}{4} F_{\mu\nu}F^{\mu\nu} -|D_\mu \Phi|^2 - V(|\Phi|),  \label{L}
\end{align}
where $F_{\mu\nu} = \partial_\mu A_\nu-\partial_\nu A_\mu$ with $A_\mu$ being the dark photon field and $D_\mu\Phi = \partial_\mu\Phi- ig A_\mu\Phi$ with $\Phi$ being the Higgs field in association with the dark U(1) gauge symmetry and $g$ is the gauge coupling constant. 
The potential $V$ is chosen as $V = \frac{\lambda}{4}\left(|\Phi|^2-v^2\right)^2$ with the Higgs self coupling constant $\lambda$ and the vacuum expectation value $v$. After the symmetry breaking, the dark photon obtains a mass of $m=\sqrt{2} gv$.

In this model, cosmic string networks are formed in the universe if the symmetry breaking happens after inflation. 
Even in the case of gauged U(1) symmetry, the actual string dynamics is almost the same as global strings for small enough gauge coupling $g$~\cite{Kitajima:2022lre}.
The dynamics of global strings, often in the context of axion, have been studied in Refs.~\cite{Davis:1985pt,Davis:1986xc,Vilenkin:1986ku,Harari:1987ht,Davis:1989nj,Dabholkar:1989ju,Hagmann:1990mj,Battye:1993jv,Battye:1994au,Yamaguchi:1998gx,Yamaguchi:1999yp,Yamaguchi:1999dy,Hagmann:2000ja,Hiramatsu:2010yu,Hiramatsu:2012gg,Fleury:2015aca,Klaer:2017qhr,Gorghetto:2018myk,Kawasaki:2018bzv,Buschmann:2019icd,Hindmarsh:2019csc,Klaer:2019fxc,Gorghetto:2020qws,Hindmarsh:2021vih,Buschmann:2021sdq,Blanco-Pillado:2022axf}.
The distribution of cosmic string networks approaches to the so-called scaling regime, in which there are $\xi\sim \mathcal O(1)$ number of infinite strings in one Hubble volume. To maintain the scaling, the network continuously should produce small string loops in each Hubble time through reconnections. 
Recently, violation of the scaling law has been reported based on highly-resolved long-term field theoretic simulations \cite{Gorghetto:2018myk,Kawasaki:2018bzv,Gorghetto:2020qws}. Numerical results suggest that $\xi$ is not constant but rather evolves logarithmically with time. Physical interpretation of such logarithmic time dependence is still unclear but it can be partially explained by the logarithmic growth of the string tension with time, i.e. $\mu \propto \log(t/\delta)$ with $\delta$ being the string width. It implies that the energy loss rate of the (infinite) string network decreases inversely proportional to $\log(t/\delta)$ due to the emission of axions/dark photons and thus the string density increases accordingly \cite{Kawasaki:2018bzv}.\footnote{Another group reported that their simulation does not show the scaling violation \cite{Hindmarsh:2019csc,Hindmarsh:2021vih}. More refined simulation is necessary to confirm whether the scaling violation occurs or not.}
String loops produced in this way will shrink by emitting either dark photons or GWs.
Loops with a length $\ell$ shorter than $m^{-1}$ loses their energy by emitting dark photon within one Hubble time, while loops with $\ell \gtrsim m^{-1}$ loses their energy only through the GW emission. 
The dark photon number density produced in one Hubble time is estimated as $n \sim \xi\mu/t$ with $\mu$ being the string tension, by assuming that the typical loop length is comparable to the Hubble radius $t$. Assuming the radiation-dominated universe when the dark photon is produced, the final energy density of the dark photon is dominated by those produced around $t \sim m^{-1}$, after which loops cannot emit vector boson. The resulting dark photon abundance is estimated as~\cite{Long:2019lwl,Kitajima:2022lre}
\begin{align}
	\Omega_{\gamma'} h^2 \simeq 0.091 \bigg(\frac{\xi}{12} \bigg) \bigg(\frac{m}{10^{-13}\,{\rm eV}}\bigg)^{1/2} \bigg(\frac{v}{10^{14}\,{\rm GeV}}\bigg)^2,
\label{OmegaA}
\end{align}
in terms of the density parameter, where $\xi \simeq 0.15\log(\sqrt{\lambda}v/m)$. It can easily explain the present dark matter abundance for wide parameter ranges.
Thus dark photon produced by cosmic string loops is one of the dark matter candidates.

Cosmic strings also produce GWs that form the stochastic GW background in the present universe~\cite{Vachaspati:1984gt,Garfinkle:1987yw,Caldwell:1991jj,Caldwell:1996en,Damour:2000wa,Damour:2001bk,Damour:2004kw,Siemens:2006yp,DePies:2007bm,Olmez:2010bi,Binetruy:2012ze,Kuroyanagi:2012wm,Kuroyanagi:2012jf,Ringeval:2017eww,Cui:2017ufi,Gouttenoire:2019kij,Gorghetto:2021fsn,Chang:2021afa}.
In the scenario of light dark photon/axion, one should take account of the effect of dark photon/axion emission by the loops, which significantly suppress the GW emission efficiency for small loops. The resulting GW spectrum is very different from both the local and global string cases. A detailed analysis of GW spectrum in such a case is performed in Ref.~\cite{Kitajima:2022lre}.
The differential present GW background energy density, $d\rho_{\rm GW}/d f$ with $f$ being the present GW frequency, produced by the cosmic string loops is given by
\begin{align}
	\frac{d\rho_{\rm GW}}{d f} = \int dt \int d\ell \,G\mu^2(t)\,S\left(\frac{\ell fa_0}{a(t)}\right)\,\frac{dn_\ell(t)}{d\ln\ell} \left(\frac{a(t)}{a_0}\right)^3.
\end{align}
where the function $S(x)$ is given by
\begin{align}
	S(x) = (q-1)2^{q-1}\Gamma_{\rm GW} \frac{\theta(x-2)}{x^q}.  \label{Sx}
\end{align}
Here $\Gamma_{\rm GW} \sim 50$ and $q=4/3$ $(5/3)$ if the cusps (kinks) on the loop dominate the GW emission~\cite{Vachaspati:1984gt,Garfinkle:1987yw,Damour:2001bk,Olmez:2010bi}. 
The loop distribution function $dn_\ell(t) / d\ln\ell$ is given by
\begin{align}
	\frac{d n_{\ell}(t)}{d\ln \ell} \simeq
	\frac{\tau(\ell)}{\tau(\ell) + t_i} \frac{C \xi (t_i)}{\alpha t_i^3}\,\left(\frac{a(t_i)}{a(t)}\right)^3 \theta\left(t-\frac{\ell_i}{\alpha}\right),
\end{align}
by assuming that newly produced loops at the cosmic time $t=t_i$ has a fixed size $\ell_i = \alpha t_i$. In a numerical calculation below we tale $\alpha=0.1$. 
The loop lifetime $\tau(\ell)$ is determined by the GW emission for $\ell \gtrsim m^{-1}$ and by the vector boson emission for $\ell \lesssim m^{-1}$:
\begin{align}
	\tau(\ell) \sim \begin{cases}
		\displaystyle \frac{\ell}{\Gamma_{\rm GW} G \mu} & {\rm for~~~} \ell \gtrsim m^{-1} \\
		\displaystyle \frac{2\pi\log(\sqrt{\lambda}v/M)}{\Gamma_{\rm vec}}\ell & {\rm for~~~} \ell \lesssim m^{-1}
	\end{cases}.
\end{align}
where $\Gamma_{\rm vec}\sim 65$~\cite{Vilenkin:2000jqa,Vilenkin:1986ku,Chang:2021afa} and $M \equiv {\rm max}[m, H]$.
We will express the present GW spectrum in terms of the density parameter:
\begin{align}
	\Omega_{\rm GW} (f) = \frac{1}{\rho_{\rm crit}} \frac{d\rho_{\rm GW}(t_0)}{d \ln f},
\end{align}
where $\rho_{\rm crit}$ denotes the critical energy density of the present universe.

\begin{figure}
\begin{center}
 \includegraphics[width=13cm]{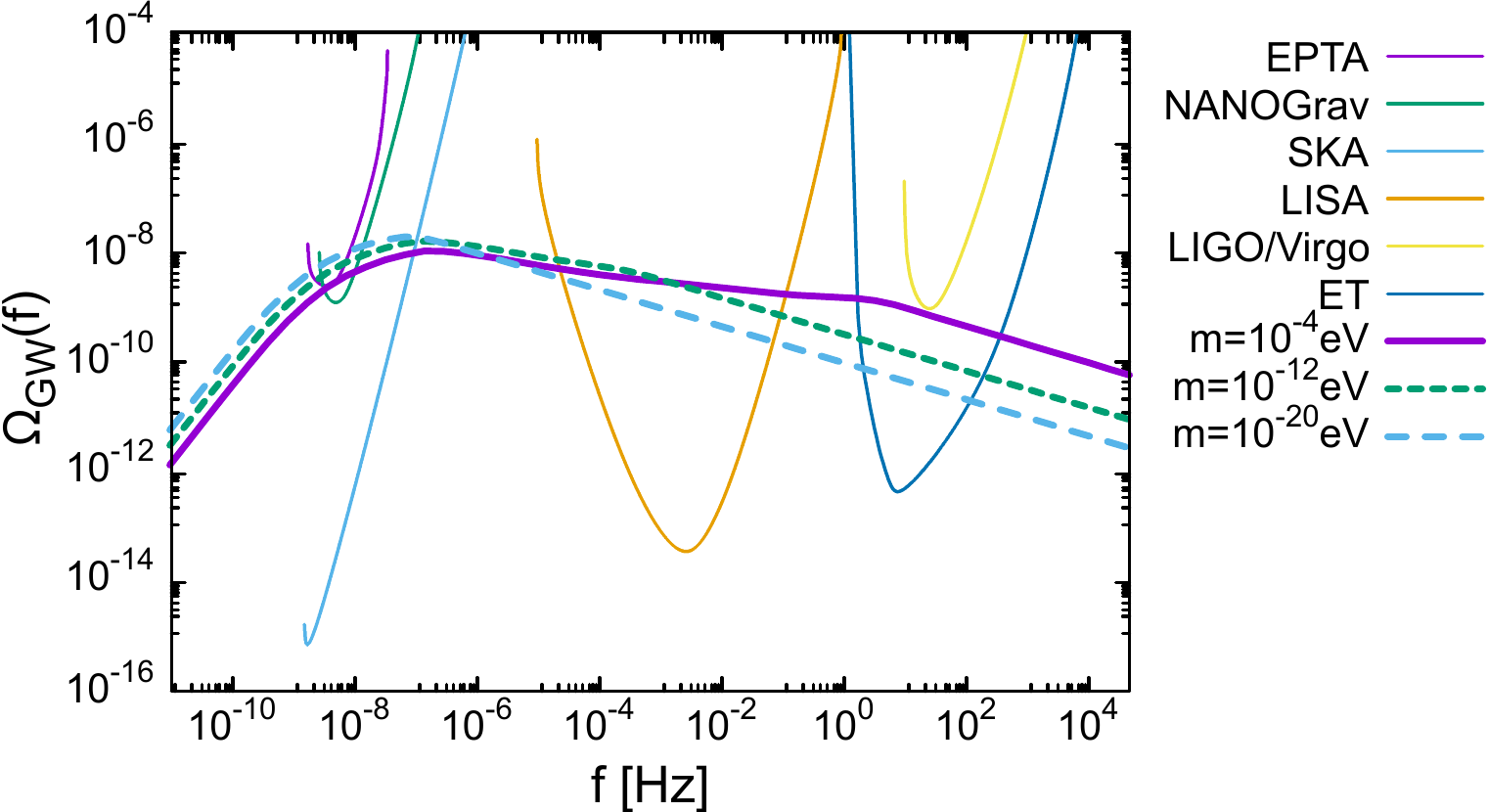}\\
 \vspace{3mm}
 \includegraphics[width=13cm]{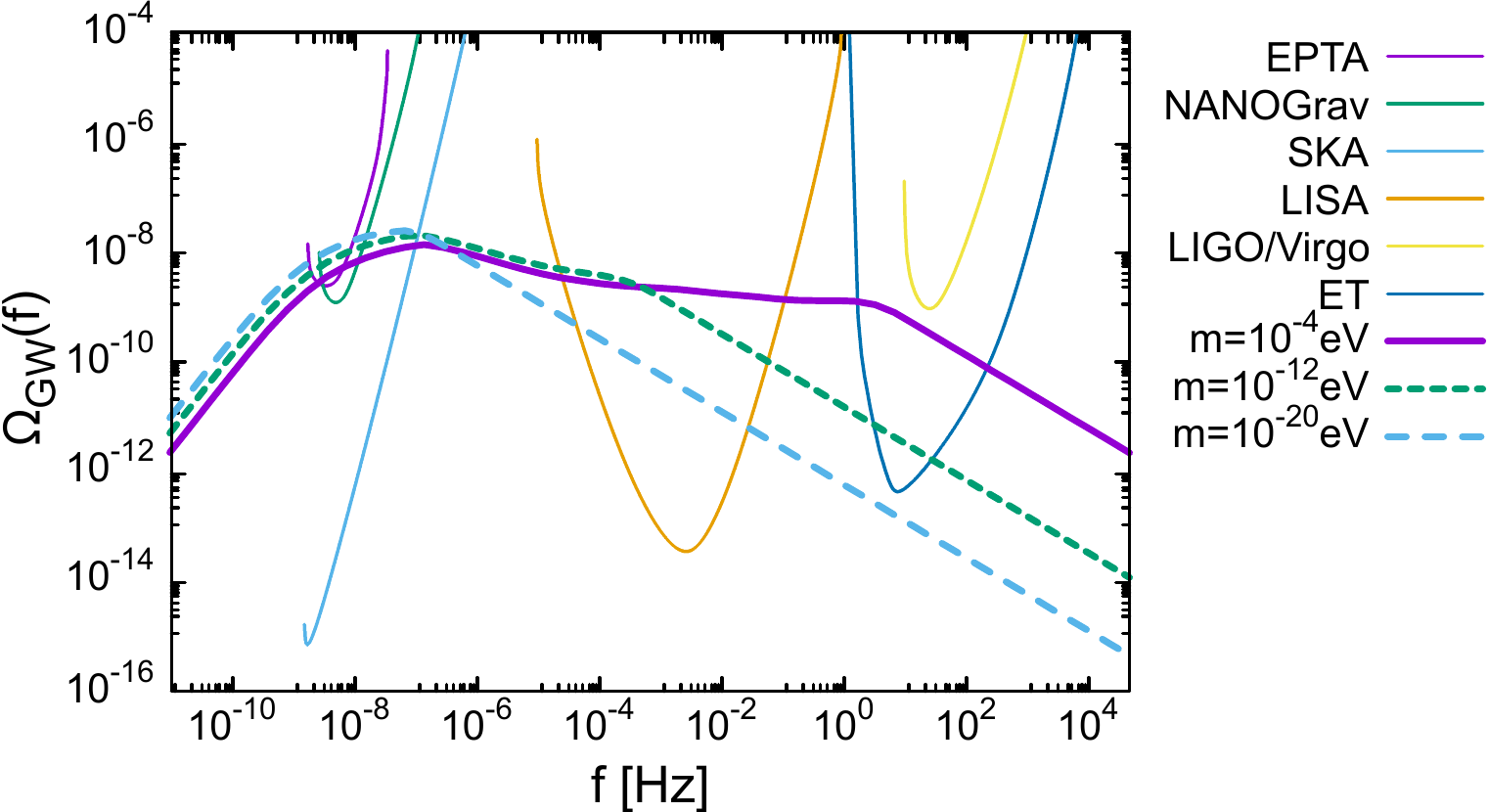}
  \end{center}
  \caption{The predicted GW spectrum from cosmic strings for $v=2\times 10^{12}\,{\rm GeV}$ and $m=(10^{-4}, 10^{-12}, 10^{-20})$\,eV. The upper (lower) figure corresponds to the case where cusps (kinks) dominate the GW emission from string loops. Note that the present dark matter abundance is explained by the dark photon produced by string loops for $m=10^{-4}$\,eV. 
  Together shown are current/future sensitivities of various experiments.}
  \label{fig:gw}
\end{figure}

Fig.~\ref{fig:gw} shows the predicted GW spectrum from cosmic strings for $v=2\times 10^{12}\,{\rm GeV}$ and $m=(10^{-4}, 10^{-12}, 10^{-20})$\,eV. We also take $\lambda=1$. The upper (lower) figure corresponds to the case where cusps (kinks) dominate the GW emission from string loops. Note that the present dark matter abundance is explained by the dark photon produced by string loops for $m=10^{-4}$\,eV (see Eq.~(\ref{OmegaA})). Together shown are future sensitivities of LIGO/Virgo~\cite{LIGOScientific:2017ikf,LIGOScientific:2014pky,VIRGO:2014yos}, Einstein Telescope (ET)~\cite{Punturo:2010zz}, LISA~\cite{Bartolo:2016ami} and Square Kilometer Array (SKA)~\cite{Janssen:2014dka}.
The current limits from pulsar timing arrays, European Pulsar Timing Array (EPTA)~\cite{vanHaasteren:2011ni} and NANOGrav 12.5 year data~\cite{NANOGrav:2020bcs} are also shown. These sensitivities are taken from Refs.~\cite{Schmitz:2020syl,Schmitz:2020}. 

It is seen that the PTA result can be explained for $v\sim 10^{12}\,{\rm GeV}$ for wide dark photon mass range $m\lesssim 10^{-4}\,{\rm eV}$. Heavier dark photon predicts too much dark matter abundance unless thermal history in the early universe is modified. The GW amplitude at the PTA frequency range is almost independent of the dark photon mass up to the small logarithmic dependence of $\mu$ and $\xi$. On the other hand, the GW amplitude at the LIGO/Virgo frequency range significantly depends on the dark photon mass. Interestingly, only for the case of $m\sim 10^{-4}\,{\rm eV}$, the GW is detectable in future LIGO/Virgo detectors and it is exactly the dark photon mass that explains the present dark matter abundance. For lower dark photon mass, GW detection at the LIGO/Virgo may not be possible, but they are detectable at future space laser interferometers, LISA or DECIGO, and also ground-based detector (ET).
In any case, it is the high frequency tail of the GW spectrum produced by cusps on the loop~(\ref{Sx}) that might be detectable at the LIGO/Virgo. However, one should note that the cusp formation on the string loops may not be guaranteed so far for the case of (nearly) global strings. On the other hand, kinks are formed in the process of string reconnection, and hence their existence on the loops may be more promising. For the case of kink-dominated GW emission, as shown in the lower panel of Fig.~\ref{fig:gw}, the GW amplitude is much suppressed for higher frequency.
Therefore, by combining the information of the pulsar timing arrays and future space/ground-based detectors, we can obtain information on the dark photon mass and the fundamental nature of cosmic string loops. 

\begin{figure}[tp]
\centering
\includegraphics [width = 13cm, clip]{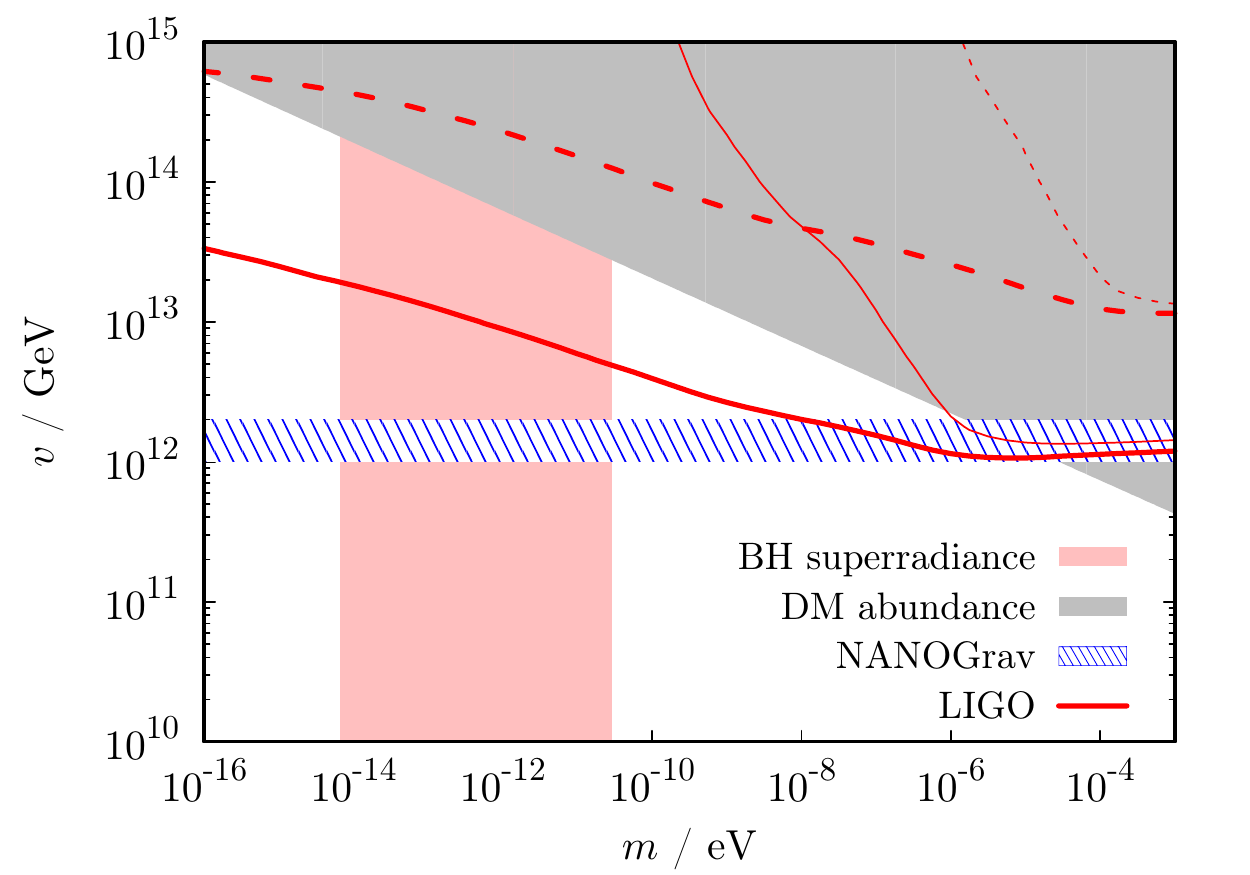}
\caption{
Constraint on the model parameter. NANOGrav result is shown by the blue hached region (factor 2 uncertainty is assumed by hand). Future designed sensitivity of LIGO/Virgo experiment is shown by the red line. Thick and thin lines represent respectively the cusp and kink contributions. Dashed-red line shows the current sensitivity.
}
\label{fig:constraint}
\end{figure}

Fig.~\ref{fig:constraint} shows the constraint on the model parameter. The gray- and orange-shaded regions are ruled out from the dark matter abundance and the black hole superradiance~\cite{Cardoso:2018tly} respectively.
It should be emphasized that PTA result (blue hached region) implies $m\sim 10^{-5}$-$10^{-6}$ eV to explain the dark matter abundance by the dark photon matter and it can be tested by future~LIGO/Virgo~\cite{LIGOScientific:2017ikf,LIGOScientific:2014pky,VIRGO:2014yos}, as shown by a red solid line. Region above the red dashed line is excluded by the current LIGO/Virgo constraint~\cite{Schmitz:2020syl}.

Let us comment on the difference between the cosmic string scenario discussed in Ref.~\cite{NANOGrav:2023hvm} and ours. For the case of stable local strings, by fixing the string tension $\mu$ so that it reproduces the PTA signal, the peak frequency is also fixed: $f_{\rm peak} \sim (\Gamma_{\rm GW} G\mu t_0)^{-1}$ and $\Omega_{\rm GW}(f_{\rm peak}) \sim G\mu$. The spectral index $n$ of $\Omega_{\rm GW}$, defined by $n=d\ln\Omega_{\rm GW}(f)/d\ln f$, is much softer than the observationally favored value. In our case, the cosmic strings are rather close to global strings, in which case the typical number of strings in one Hubble volume, $\xi$, is likely to be logarithmically enhanced~\cite{Hiramatsu:2012gg,Fleury:2015aca,Klaer:2017qhr,Gorghetto:2018myk,Kawasaki:2018bzv,Buschmann:2019icd,Klaer:2019fxc,Gorghetto:2020qws,Buschmann:2021sdq,Blanco-Pillado:2022axf,Kitajima:2022lre}, although the discussion is not settled yet~\cite{Hindmarsh:2019csc,Hindmarsh:2021vih}.  
Thus the GW amplitude is enhance by this factor, which makes it possible that the hard part of the GW spectrum reaches to the PTA level. In this sense, our scenario is phenomenologically similar to the cosmic superstrings with small reconnection probability mentioned in Ref.~\cite{NANOGrav:2023hvm}, which shows good agreement with the observation. For the superstring case, the current result from the LIGO/Virgo experiment severely constrains the parameter region unless some modified cosmological history is assumed, while in our case the GW spectrum is naturally suppressed at the LIGO/Virgo frequency. This is because small loops decays to dark photons and the efficiency for the GW emission is suppressed at high frequency.

\begin{figure}
\begin{center}
 \includegraphics[width=10cm]{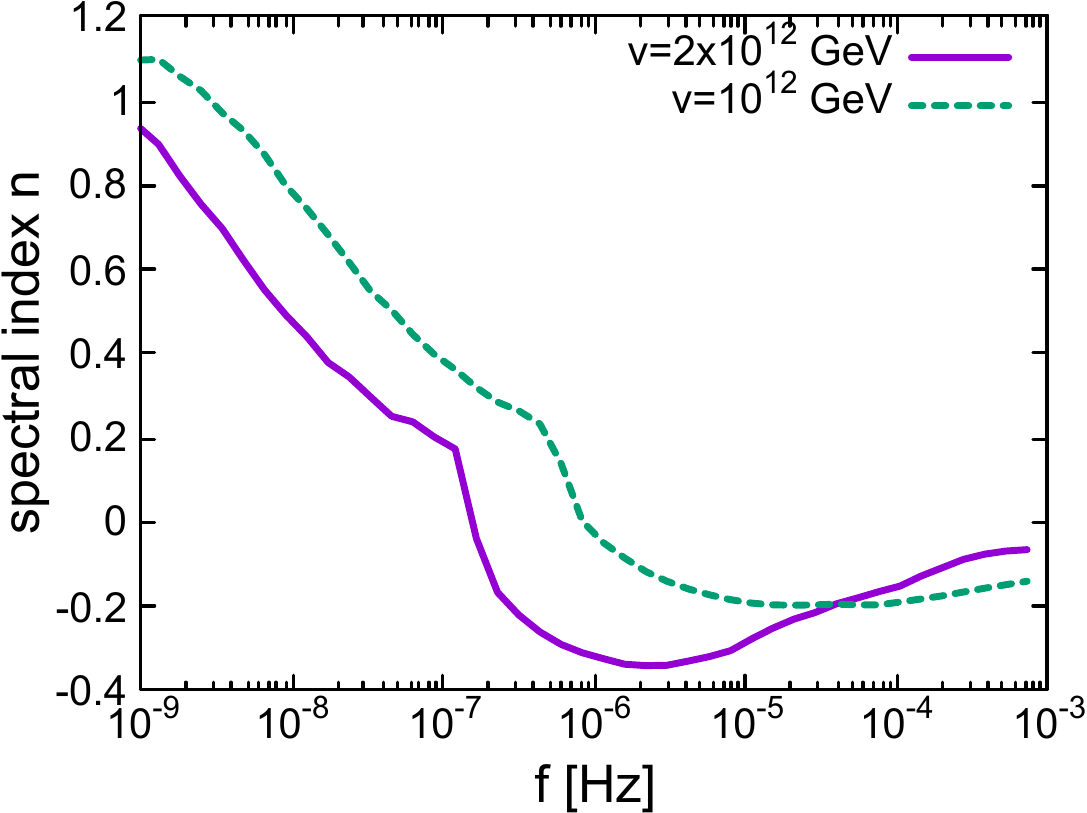}
  \end{center}
  \caption{The spectral index $n$ of $\Omega_{\rm GW}(f)$ for $v=2\times 10^{12}\,{\rm GeV}$ and $v=10^{12}\,{\rm GeV}$.}
  \label{fig:n}
\end{figure}

It should be noticed that theoretical prediction of the GW spectrum suffers from various uncertainties due to incomplete understandings of long term cosmic string dynamics. One should consider that our estimation may have at least $\mathcal O(1)$ numerical uncertainty. 
Although we have used $v=2\times 10^{12}\,{\rm GeV}$ as a reference value, $v$ may have a bit large or smaller values to explain the PTA result taking account of theoretical uncertainty.
Fig.~\ref{fig:n} shows the spectral index $n$ in our scenario for $v=2\times 10^{12}\,{\rm GeV}$ and $v=10^{12}\,{\rm GeV}$. 
Around the PTA frequency range it is around $0.5$--$0.8$. Accidentally it is close to the case of supermassive blackhole binaries~\cite{NANOGrav:2023gor}, which typically predicts $n=2/3$.
Therefore our model fits the data at the same level as the supermassive blackhole binaries.

To summarize, the PTA result for the detection of GWs may be explained by the cosmic strings interacting with light dark photon. The dark photon mass is predicted to be $m \sim 10^{-6}$--$10^{-4}$\,eV if the dark photon produced by strings accounts for the present dark matter density of the universe. Note that GWs with higher frequency band (e.g. $\mathcal O(10\,{\rm Hz})$) emitted by the same string network is within the sensitivity of future LIGO/Virgo observation, while the current constraint is evaded.
Dark photon with this mass range may be proved by several experiments~\cite{Caputo:2021eaa} such as haloscope, if the dark photon has a small but finite kinetic mixing with the Standard Model photon.
Such a dark photon dark matter scenario is likely to predict observable level of GW amplitude at the LIGO/Virgo frequency range. 
On the other hand, we also need more robust theoretical understandings for the long term cosmological dynamics of strings in order to predict the GW spectrum precisely.

\section*{Acknowledgment}

This work was supported by JSPS KAKENHI Grant Nos. 19H01894 (N.K.), 20H01894 (N.K.), 20H05851 (N.K.), 21H01078 (N.K.), 21KK0050 (N.K.).
This work was supported by World Premier International Research Center Initiative (WPI), MEXT, Japan.

\bibliographystyle{utphys}
\bibliography{ref}

\end{document}